\documentclass[%
 preprint,
 amsmath,amssymb,
 aps,
 pre,
showkeys
]{revtex4-2}

\usepackage{graphicx}
\usepackage{dcolumn}
\usepackage{bm}
\usepackage{hyperref}
\usepackage{xcolor}
\usepackage{ctable}

\begin{document}

\title{Critical temperature and thermodynamic properties of a homogeneous dilute weakly interacting Bose gas within the improved Hartree-Fock approximation at finite temperature}

\author{Nguyen Van Thu and Pham Duy Thanh}
\affiliation{Department of Physics, Hanoi Pedagogical University 2, Hanoi 100000, Vietnam}
\email[]{nvthu@live.com}


\begin{abstract}
By means of Cornwall–Jackiw–Tomboulis effective action approach we investigate a homogeneous dilute weakly interacting Bose gas at finite temperature in vicinity of critical region. A longstanding debate, the shift of critical temperature, is considered and obtained in the universal form $\Delta T_C/T_C^{(0)}=c.(n_0a_s^3)^a$ with constants $c$ and $a$. The non-condensate fraction is contributed by quantum fluctuations as well as thermal exitations and can be expressed in sum of three terms. These terms correspond to the quantum fluctuations, thermal fluctuations and both. Indeed, the specific heat capacity and critical exponents are calculated and in excellent agreement with those in previous works and experimental data.
\end{abstract}


\maketitle

\section{Introduction\label{sec1}}

It is well-known that a system of Bose gas undergoes a condensate phase transition when its temperature is lowed to the critical temperature \cite{Bose,Einstein}.  The result is that a number of atoms in the Bose gas will be condensed and formed a Bose-Einstein condensate (BEC). The criterion for the BEC to be formed in macroscopic scale is \cite{Pethick},
\begin{eqnarray}
n_0\lambda_B^3\leq\zeta(3/2),\label{criterion}
\end{eqnarray}
in which $n_0$ is total particle density, which is defined as the number of particle in a unit of volume, $\zeta(x)=\sum_{n=1}^\infty 1/n^x$ is zeta function. The de Broglie wave length associated with a particle with mass of $m$ at temperature $T$ reads
\begin{eqnarray}
\lambda_B=\left(\frac{2\pi\hbar^2}{mk_BT}\right)^{1/2},\label{wave}
\end{eqnarray}
with $\hbar$ and $k_B$ being the reduced Planck and Boltzmann constants, respectively. In case of the ideal Bose gas, Eq. (\ref{criterion}) gives the critical temperature
\begin{eqnarray}
T_C^{(0)}=\frac{2\pi\hbar^2}{mk_B}\left[\frac{n_0}{\zeta(3/2)}\right]^{2/3},\label{Tideal}
\end{eqnarray}
which is presented in many relevant textbooks \cite{Huang1987,Pethick}.

For a weakly interacting Bose gas, the interatomic interaction is assumed to be pairwise and described by repulsive contact pseudo-potential (Lieb–Liniger model) \cite{Lieb1963,Lieb1963a}. The strength of this interaction is determined by a coupling constant $g=4\pi\hbar^2a_s/m$ with $a_s$ the scattering length of the $s$-wave. Owing to the repulsive interaction among bosonic atoms, the critical temperature is shifted in compared with that of the ideal Bose gas
\begin{eqnarray}
\frac{\Delta T_C}{T_C^{(0)}}=c\alpha_s^{1/3},\label{shift}
\end{eqnarray}
where $\alpha_s=n_0a_s^3$ is called the gas parameter and $\Delta T_C\equiv T_C-T_C^{(0)}$. The condition of diluteness of the Bose gas is $\alpha_s\ll1$ \cite{Andersen}, i.e. the $s$-wave scattering $a_s$ length is very small in compared with the interparticle distance $n_0^{-1/3}$. Value and characteristics of the constant $c$ have attracted the attention of many physicists to compute, such as \cite{Toyoda1982, Wilkens2000, Ledowski2004, Davis2003, Arnold2001, Baym2000, SouzaCruz2001, Stoof1992}. However, some results showed negative $c$ whereas others obtained the positive $c$. This issue still remains a matter of controversy.

Below the critical temperature, the lower the temperature is, the more condensate particles there are. For the ideal Bose gas, when temperature is smaller than the critical temperature $T_C^{(0)}$, the condensate density is \cite{Pitaevskii},
\begin{eqnarray}
n=n_0\left[1-\left(\frac{T}{T_C^{(0)}}\right)^{3/2}\right].\label{dens}
\end{eqnarray}
For the dilute weakly interacting Bose gas, the non-condensate fraction has been calculated in several different approximations, for instance, Hartree-Fock, Hartree-Fock Bogoliubov, Popov approximations. A typical result is \cite{Shi1998},
\begin{eqnarray}
\frac{n_{\rm ex}}{n_0}=\frac{8}{3\sqrt{\pi}}\alpha_s^{1/2}+\frac{m}{12\hbar^2cn_0}(k_BT)^2,\label{Shi}
\end{eqnarray}
where $c$ is speed of sound in the Bose gas.

Theoretically, at zero temperature all of the atoms will stay in the ground state \cite{Pitaevskii}. In this situation, the ground state is described by a wave function, which can be found by solving the Gross-Pitaevskii (GP) equation \cite{Gross,Pitaevskii1}. However, due to interaction-induced quantum fluctuations in the BEC, some particles with nonzero momentum reside in excited states instead of the ground state, even at zero absolute temperature \cite{Pethick}. These particles are pushed out of the condensate-- the phenomenon of quantum depletion. The number of atoms in the remaining condensate fraction due to quantum depletion was first studied by N. N. Bogoliubov \cite{Bogolyubov} in 1947. Whose study was just limited to the order $1/2$ in the gas parameter by using the second quantization formalism. The main idea is based on a quantum description, where the particle operators are transformed into quasi-particle operators, yielding an explicit diagonalization of the quantum Hamiltonian. In 1997, using the Bogoliubov theory and the semiclassical approach, the authors of Ref. \cite{Dalfovo} investigated the quantum depletion in a Bose gas confined by a harmonic trap. In the case of a homogeneous Bose gas their result corresponded exactly with Bogoliubov's result. Within the GP theory, the condensed fraction was reproduced by S. Stringari \cite{Stringari}. Using an effective theory called simplified approach, Carlen {\it et. al.} \cite{Elliott} verified the Bogoliubov's result for the condensed fraction of the BEC at low density.

Beyond these theorems, a powerful tool to investigate the Bose gas is Cornwall-Jackiw-Tomboulis (CJT) effective action approach \cite{CJT}. Recently, this method has been invoked to consider effect from the quantum fluctuations in the dilute BEC at zero temperature to non-condensate fraction \cite{VanThu2022,Thu2023} as well as the Casimir effect in a series of our previous papers, such as \cite{Thu,VanThu2017,VanThu2022a,VanThu2021}. At finite temperature, CJT effective action approach was employed to study the thermodynamics of a weakly interacting Bose gas \cite{Haugset1998}. However, calculations in this work were only dealt in the one-loop approximation. As a consequence, an important quantity-the critical temperature-of the weakly interacting Bose gas was pointed out the same as the one associated with the non-interacting Bose gas. The main purpose of the present paper is to investigate the critical temperature and thermodynamic properties of a dilute homogeneous weakly interacting Bose gas in vicinity of critical point within IHF approximation, in which the two-loop diagrams are taken into account.

This paper is organised as follows. In Section \ref{sec:2} we calculate the gap and Schwinger-Dyson (SD) equations for a single weakly interacting Bose gas in the improved Hartree-Fock (IHF) approximation by first recapitulating the regular Hartree-Fock (HF) method and then calculating these expressions for the CJT effective potential with symmetry-restoring terms. The critical temperature and other thermodynamic quantities are investigated in Section \ref{sec:3}. Finally, we present the conclusions and a future outlook in Section \ref{sec:4}.

\section{The equations of state in the IHF approximation}\label{sec:2}

In this Section, we will establish the equations of state for a Bose gas, which consists of the gap and Schwinger-Dyson (SD) equations. To do so, the first our task is to derive the CJT effective potential \cite{CJT} in the IHF approximation. We set the stage for our calculations by starting with a dilute Bose gas described by the following Lagrangian density without any external field \cite{Pethick},
\begin{equation}
{\cal L}=\psi^*\left(-i\hbar\frac{\partial}{\partial t}-\frac{\hbar^2}{2m}\nabla^2\right)\psi-\mu\left|\psi\right|^2+\frac{g}{2}\left|\psi\right|^4,\label{eq:1}
\end{equation}
wherein $\mu$ is the chemical potential. The field operator $\psi(\vec{r},t)$ depends on both the coordinate $\vec{r}$ and time $t$. According to Ref. \cite{Pethick}, the interaction potential between the atoms can be chosen as the hard-sphere model. In this model, the strength of the interatomic interaction is determined by the coupling constant $g=4\pi\hbar^2a_s/m$, which is expressed in terms of the $s$-wave scattering length $a_s$ by making use of the Born approximation. The thermodynamic stability requires that $g>0$, i.e., the boson interactions are repulsive with positive $s$-wave scattering length. Indeed, from now on, we consider the system in the grand canonical ensemble, i.e the total particle density is fixed.

In tree-approximation, the expectation value $\psi_0$ of the field operator is independent of both time and coordinate, therefore the GP potential is then taken from \eqref{eq:1}
\begin{equation}
V_{GP}=-\mu\psi_0^2+\frac{g}{2}\psi_0^4.\label{eq:VGP}
\end{equation}
Furthermore, the non-macroscopic part of the condensate moves as a whole so that the expectation value $\psi_0$ is real and plays the role of the order parameter. The square of the order parameter $\psi_0^2$ is defined as the condensate density. Minimizing the potential \eqref{eq:VGP} with respect to the order parameter, one arrives at the gap equation
\begin{equation}
\psi_0(-\mu+g\psi_0^2)=0,\label{eq:gaptree}
\end{equation}
and hence, in the broken phase the non-trivial solution is read
\begin{equation}
\psi_0^2=\frac{\mu}{g}.\label{eq:psi0}
\end{equation}
In order to proceed within the framework of the HF approximation, the complex field operator $\psi$ should first be decomposed in terms of the order parameter $\psi_0$ and two real fields $\psi_1$ and $\psi_2$, which are associated with quantum fluctuations of the field \cite{Andersen}, i.e.,
\begin{equation}
\psi\rightarrow \psi_0+\frac{1}{\sqrt{2}}(\psi_1+i\psi_2).\label{eq:shift}
\end{equation}
Plugging equation \eqref{eq:shift} into \eqref{eq:1}, the Lagrangian density is readily devided into three parts, namely

- The free Lagrangian density ${\cal L}_0$, which gives us the free propagator.

- The tree Lagrangian density is associated with the tree-approximation. This part allows us to find the propagator (or Green's function) in the tree-approximation
\begin{equation}
D_0(k)=\frac{1}{\omega_n^2+E_{\text{(tree)}}^2(k)}\left(
              \begin{array}{cc}
                \frac{\hbar^2k^2}{2m}-\mu+g\psi_0^2 & \omega_n \\
                -\omega_n &  \frac{\hbar^2k^2}{2m}-\mu+3g\psi_0^2\\
              \end{array}
            \right),\label{eq:protree}
\end{equation}
where $\omega_n$ is the $n$th Matsubara frequency for bosons. The latter is defined as $\omega_n=2\pi n k_BT$ where $n\in{\mathbb{Z}}$. The $E_{\rm{(tree)}}(k)$ is the energy spectrum of the elementary excitation, which is determined by examining poles of the Green's function \cite{Andersen,Negele}. Together with the gap equation \eqref{eq:psi0}, the result is
\begin{equation}
E_{\rm{(tree)}}(k)=\sqrt{\frac{\hbar^2k^2}{2m}\left(\frac{\hbar^2k^2}{2m}+2g\psi_0^2\right)}.\label{dispertree}
\end{equation}
The spectrum Eq. (\ref{dispertree}) was first achieved by Bogoliubov \cite{Bogolyubov}. For small momenta, this equation is gapless, which represents the phonon excitations. This is due on the Goldstone theorem \cite{Goldstone}, which requires that the dispersion relation is linear in the wave vactor whenever a continuous symmetry $U(1)$ is spontaneously broken \cite{Anderson2018}. In the physics of BEC, this theorem is referred to as the Hugenholtz-Pines theorem \cite{Hugenholtz} at zero temperature and a more general proof for all value of temperature was given by Hohenberg  and Martin \cite{Hohenberg}. More discussion about the gapless excitation spectrum for the BEC at finite temperature were mentioned in Refs. \cite{Hutchinson,Griffin}.

- The last one is the interaction Lagrangian density, which describes the interaction between the two real fields of fluctuations. In the HF approximation, keeping up to the cubic and quartic terms this Lagrangian density is obtained
\begin{equation}
{\cal L}_{\rm int}=\frac{g}{2}\psi_0\psi_1(\psi_1^2+\psi_2^2)+\frac{g}{8}(\psi_1^2+\psi_2^2)^2.\label{eq:Lint}
\end{equation}

To continue our discussion, we introduce the CJT effective potential in the HF approximation that can be constructed from the interaction Lagrangian density \eqref{eq:Lint} in the manner that was pointed out in \cite{Thu},
\begin{equation}
\begin{split}
V_T^{\text{(CJT)}} =&-\mu\psi_0^2 +\frac{g}{2}\psi_0^4+\frac{1}{2}\int_T \mbox{tr}\left[\ln G^{-1}(k)+D_0^{-1}(k)G(k)-{1\!\!1}\right]\\
&+\frac{3g}{8}(P_{11}^2+P_{22}^2)+\frac{g}{4}P_{11}P_{22}\, ,\label{eq:VHF}
\end{split}
\end{equation}
for which the functions $P_{11}$ and $P_{22}$ are
\begin{subequations}
    \begin{equation}
        \label{eq:P11}
        P_{11}=\int_T G_{11}(k)
    \end{equation}
    \begin{equation}
        \label{eq:P22}
        P_{22}=\int_T G_{22}(k)
    \end{equation}\label{P}
\end{subequations}
The Matsubara integrals in these expressions are defined as follows
\begin{equation}
    \label{eq:fk}
    \int_T f(k)=k_BT\sum_{n=-\infty}^{+\infty}\int\frac{d^3\vec{k}}{(2\pi)^3}f(\omega_n,\vec{k})\, .
\end{equation}
Here $G(k)$ is the propagator or Green's function in the HF approximation, which can be obtained by minimizing the CJT effective potential \eqref{eq:VHF} with respect to the elements of the propagator. Performing these calculations result in the following expression for the inverse propagator
\begin{equation}
G^{-1}(k)=D_0^{-1}(k)+\Pi,\label{eq:r11}
\end{equation}
in which
\begin{equation}
\Pi=\left(
              \begin{array}{cc}
                \Pi_1& 0 \\
                0 & \Pi_2\\
              \end{array}
            \right),\label{eq:r12}
\end{equation}
with the matrix entries $\Pi_1$ and $\Pi_1$ being the self-energies that can readily be constructed from \eqref{eq:P11} and \eqref{eq:P22}, i.e.,
\begin{subequations}
    \begin{equation}
        \label{eq:r13}
        \Pi_1=\frac{3g}{2}P_{11}+\frac{g}{2}P_{22}
    \end{equation}
    \begin{equation}
        \label{eq:r14}
        \Pi_2=\frac{g}{2}P_{11}+\frac{3g}{2}P_{22}\, .
    \end{equation}
\end{subequations}
The gap equation in the HF approximation can now be found by minimizing the CJT effective potential \eqref{eq:VHF} with respect to the order parameter $\psi_0$, i.e.,
\begin{equation}
-\mu+g\psi_0^2+\Pi_1=0.\label{eq:r15}
\end{equation}
Combining equations \eqref{eq:r11}-\eqref{eq:r15}, one has the propagator in the HF approximation
\begin{equation}
\label{eq:inverse_G_CJT}
G(k)=\frac{1}{\omega_n^2+E_{\text{(HF)}}^2(k)}\left(
              \begin{array}{cc}
                \frac{\hbar^2k^2}{2m}-\mu+g\psi_0^2+\Pi_2 & -\omega_n \\
                \omega_n &  \frac{\hbar^2k^2}{2m}-\mu+3g\psi_0^2+\Pi_1\\
              \end{array}
            \right),
\end{equation}
and consequently the dispersion relation in this approximation is
    \begin{equation}
        E_{\text{(HF)}}(k)=\sqrt{\left(\frac{\hbar^2k^2}{2m}-\mu+3g\psi_0^2+\Pi_1\right)\left(\frac{\hbar^2k^2}{2m}-\mu+g\psi_0^2+\Pi_2\right)}.\label{eq:r16}
    \end{equation}
Equations \eqref{eq:r15} and \eqref{eq:r16} show a non-gapless spectrum, which means that the Goldstone theorem is violated in the case of spontaneously broken symmetry within the HF approximation. We now recall a special emphasis on enforcing the consequences of the Goldstone theorem. To do so, we employ the method developed in \cite{Ivanov}. In this way, a phenomenological symmetry-restoring  correction term $\Delta V$ need to be added into CJT effective potential, which has to contemporaneously satisfy three conditions:

(i) it restores the Goldstone boson in the condensed phase;

(ii) it does not change the HF equations for the mean field and;

(iii) the results in the phase of restored symmetry are not changed.\\
It is not difficult to readily check that this term has to be in from \cite{VanThu2022},
\begin{equation}
\Delta V=-\frac{g}{4}(P_{11}^2+P_{22}^2)+\frac{g}{2}P_{11}P_{22},\label{extra}
\end{equation}
Let the propagator in the IHF approximation be denoted as $D_{\text{(IHF)}}(k)$, the CJT effective potential \eqref{eq:VHF} now becomes
\begin{eqnarray}
    \label{eq:VIHF}
        \widetilde{V}_T^{\rm{(CJT)}}& =&  -\mu\psi_0^2 +\frac{g}{2}\psi_0^4+\frac{1}{2}\int_T \mbox{tr}\left[\ln D^{-1}_{\rm{(IHF)}}(k)+D_0^{-1}(k)D_{\rm{(IHF)}}(k)-{1\!\!1}\right]\nonumber\\
&&+\frac{g}{8}(P_{11}^2+P_{22}^2)+\frac{3g}{4}P_{11}P_{22}.
\end{eqnarray}
From the CJT effective potential in the IHF approximation \eqref{eq:VIHF}, one arrives at the gap equation
\begin{equation}
-\mu+g\psi_0^2+\Sigma_1=0,\label{eq:gap}
\end{equation}
and the SD equation
\begin{equation}
M=-\mu+3g\psi_0^2+\Sigma_2,\label{eq:SD}
\end{equation}
in which the self-energies $\Sigma_1$ and $\Sigma_2$ are
\begin{subequations}
    \begin{equation}
        \label{eq:sig1}
        \Sigma_1=\frac{3g}{2}P_{11}+\frac{g}{2}P_{22}
    \end{equation}
    \begin{equation}
        \label{eq:sig2}
        \Sigma_2=\frac{g}{2}P_{11}+\frac{3g}{2}P_{22}\, .
    \end{equation}
\end{subequations}

Combining equations \eqref{eq:VIHF}-\eqref{eq:sig2}, one can once again calculate the propagator, now in the IHF approximation, i.e.,
\begin{equation}
D_{\text{(IHF)}}(k)=\frac{1}{\omega_n^2+E_{\text{(IHF)}}^2(k)}\left(
              \begin{array}{lr}
                \frac{\hbar^2k^2}{2m} & \omega_n \\
                -\omega_n & \frac{\hbar^2k^2}{2m}+M \\
              \end{array}
            \right).\label{eq:proIHF}
\end{equation}
Hence, the resulting dispersion relation is
\begin{equation}
E_{\rm{(IHF)}}(k)=\sqrt{\frac{\hbar^2k^2}{2m}\left(\frac{\hbar^2k^2}{2m}+M\right)}.\label{disperIHF}
\end{equation}
Clearly, the Goldstone boson is restored in this approximation. The gap and SD equations \eqref{eq:gap} and \eqref{eq:SD}, together with the momentum integrals \eqref{tichphan} form the equations of state, which govern the variation of all quantities of the system.

To do further, we first deal with the momentum integrals in the gap and SD equations. Without loss of generality, henceforth we use the abbreviation $E(k)$ replacing for $E_{\rm (IHF)}(k)$.   In the IHF approximation, the momentum integrals are obtained from equations \eqref{P} after replacing $G(k)$ by $D(k)$. Using the following formula \cite{Schmitt},
\begin{equation}
\sum_{n=-\infty}^{+\infty}\frac{1}{\omega_n^2+E^2(k)}=\frac{1}{2k_BTE(k)}\left[1+\frac{2}{e^{E(k)/k_BT}-1}\right],
\end{equation}
one has
\begin{subequations}
\begin{eqnarray}
P_{11}&=&\int\frac{d^3\vec{k}}{(2\pi)^3}\frac{1}{2E(k)}\frac{\hbar^2k^2}{2m}+\int\frac{d^3\vec{k}}{(2\pi)^3}\frac{1}{E(k)}\frac{1}{e^{E(k)/k_BT}-1}\frac{\hbar^2k^2}{2m},\label{tichphan1}\\
P_{22}&=&\int\frac{d^3\vec{k}}{(2\pi)^3}\frac{1}{2E(k)}\left(\frac{\hbar^2k^2}{2m}+M\right)+\int\frac{d^3\vec{k}}{(2\pi)^3}\frac{1}{E(k)}\frac{1}{e^{E(k)/k_BT}-1}\left(\frac{\hbar^2k^2}{2m}+M\right).\label{tichphan2}
\end{eqnarray}\label{tichphan}
\end{subequations}
The first terms in right-hand side of (\ref{tichphan1}) and (\ref{tichphan2}) do not explicitly depend on the temperature. These integrals are ultraviolet divergent. This divergence is avoidable by means of the dimensional regularization \cite{Andersen} and thus, the integrals can be computed. The integral $I_{m,n}$ is
\begin{eqnarray}
        I_{m,n}({\cal M})&=&\int\frac{d^d\kappa}{(2\pi)^d}\frac{\kappa^{2m-n}}{(\kappa^2+{\cal M}^2)^{n/2}}\nonumber\\
        &=&\frac{\Omega_d}{(2\pi)^d}\Lambda^{2\epsilon}{\cal M}^{d+2(m-n)}\frac{\Gamma\left(\frac{d-n}{2}+m\right)\Gamma\left(n-m-\frac{d}{2}\right)}{2\Gamma\left(\frac{n}{2}\right)},\label{eq:tp}
\end{eqnarray}
where $\Gamma(x)$ is the gamma function, $\Omega_d=2\pi^{d/2}/\Gamma(d/2)$ is the surface area of a $d-$dimensional sphere and $\Lambda$ is a renormalization scale that ensures the integral has the correct canonical dimension. This number is usually absorbed into the measure, hence it will not appear in the results.  Applying \eqref{eq:tp} to \eqref{tichphan} with $d=3$, one finds
\begin{subequations}
    \begin{eqnarray}
    P_{110}&\equiv&\int\frac{d^3\vec{k}}{(2\pi)^3}\frac{1}{2E(k)}\frac{\hbar^2k^2}{2m}=\frac{(2m)^{3/2}M^{3/2}}{6\pi^2\hbar^3},\label{tichphan01}\\
    P_{220}&\equiv&\int\frac{d^3\vec{k}}{(2\pi)^3}\frac{1}{2E(k)}\left(\frac{\hbar^2k^2}{2m}+M\right)=-\frac{(2m)^{3/2}M^{3/2}}{12\pi^2\hbar^3}.\label{tichphan20}
    \end{eqnarray}\label{tichphan0}
\end{subequations}

Generally, the temperature-dependent integrals (\ref{tichphan}) cannot be calculated analytically. We can solve by either numerical method or in some limit cases.

\section{The critical temperature and thermodynamic properties of a homogeneous dilute Bose gas\label{sec:3}}

Let us now investigate the critical temperature and thermodynamic properties of the dilute weakly interacting Bose gas in the IHF approximation. To this end, we first consider the pressure. The pressure is defined as the negative of the CJT effective potential \eqref{eq:VIHF} at the minimum, i.e. satisfying both the gap and SD equations
\begin{equation}
{\cal P}=-\widetilde{V}_T\bigg|_{\rm{minimum}}\equiv -\widetilde{{\cal V}}_T^{\rm{(CJT)}}.\label{eq:press}
\end{equation}
Now, substituting equations \eqref{eq:gap} and \eqref{eq:SD} into \eqref{eq:VIHF}, one has
\begin{eqnarray}
    \label{eq:V1}
        \widetilde{{\cal V}}_T^{\rm{(CJT)}}&=&-\mu\psi_0^2+\frac{g}{2}\psi_0^4+\frac{1}{2}\int_T \mbox{tr}\ln D^{-1}_{\rm{(IHF)}}(k)\nonumber\\
        &&+\frac{1}{2}(3g\psi_0^2-\mu-M)P_{11}\nonumber
    +\frac{1}{2}(g\psi_0^2-\mu)P_{22}\\
    &&+\frac{g}{8}(P_{11}^2+P_{22}^2)+\frac{3g}{4}P_{11}P_{22}.
\end{eqnarray}
The chemical potential is defined as first derivative of the pressure with respect to the condensate density \cite{Landau1980}
\begin{eqnarray}
\mu=\frac{\partial \cal P}{\partial n}.\label{chemical}
\end{eqnarray}
Plugging (\ref{eq:V1}) into (\ref{chemical}) one can find the chemical potential beyond the mean-field theory
\begin{equation}
\mu=gn_0+gP_{11}.\label{eq:chemical}
\end{equation}
Combining (\ref{eq:chemical}), (\ref{eq:gap}) and (\ref{eq:SD}) leads to the gap and SD equations in higher precision
\begin{eqnarray}
&&-1+x+\frac{1}{2n_0}(P_{11}+P_{22})=0,\nonumber\\
&&{\cal M}=-1+3x-\frac{1}{2n_0}(P_{11}-3P_{22}),\label{eqnew}
\end{eqnarray}
in which we defined ${\cal M}=M/gn_0$ and $x=n_c/n_0$ is the dimensionless effective mass and condensate fraction, respectively.

In order to solve Eqs. (\ref{eqnew}) for two unknowns $x$ and ${\cal M}$ we must calculate the temperature-dependent part of the momentum integrals (\ref{tichphan}). As already mentioned above, this part cannot be analytically calculated. With the aim finding the critical temperature, we consider the high-temperature limit, i.e the temperature is not much lower than the critical temperature. In this limit $M/k_BT\ll1$ and in lowest order approximation, the temperature-dependent parts (\ref{tichphan}) of the momentum integrals are
\begin{eqnarray}
P_{11T}=P_{22T}=\frac{m^{3/2}\zeta(3/2)}{2\sqrt{2}\pi^{3/2}\hbar^3}(k_BT)^{3/2}.\label{tichphanTT}
\end{eqnarray}
In sum, at high temperature, the momentum integrals (\ref{tichphan}) can be expressed
\begin{eqnarray}
P_{11}&=&\frac{(2m)^{3/2}M^{3/2}}{6\pi^2\hbar^3}+\frac{m^{3/2}\zeta(3/2)}{2\sqrt{2}\pi^{3/2}\hbar^3}(k_BT)^{3/2},\nonumber\\
P_{22}&=&-\frac{(2m)^{3/2}M^{3/2}}{12\pi^2\hbar^3}+\frac{m^{3/2}\zeta(3/2)}{2\sqrt{2}\pi^{3/2}\hbar^3}(k_BT)^{3/2}.\label{chottichphan}
\end{eqnarray}

\begin{figure}
  \includegraphics[width = 0.6\linewidth]{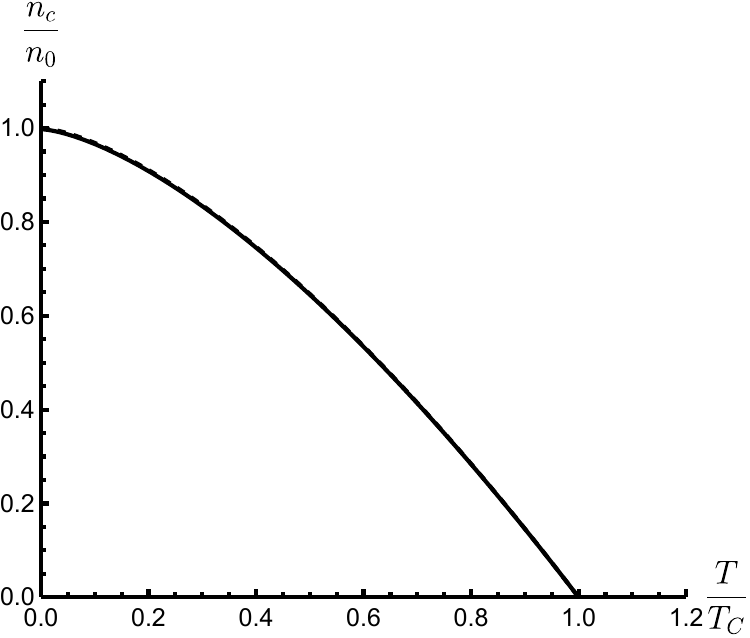}
\caption{(Color online) The condensate fraction as a function of the reduced temperature $T/T_C$.}
\label{fig:f1}       
\end{figure}

Inserting (\ref{chottichphan}) into (\ref{eqnew}) one can write the gap and SD equations in dimensionless form
\begin{eqnarray}
&&-1+x+\frac{2\sqrt{2}}{3\sqrt{\pi}}{\cal M}^{3/2}\alpha_s^{1/2}+\frac{m^{3/2}\zeta(3/2)}{2\sqrt{2}\pi^{3/2}\hbar^3n_0}(k_BT)^{3/2}=0,\nonumber\\
&&{\cal M}=-1+3x-\frac{10\sqrt{2}}{3\sqrt{\pi}}{\cal M}^{3/2}\alpha_s^{1/2}+\frac{m^{3/2}\zeta(3/2)}{2\sqrt{2}\pi^{3/2}\hbar^3n_0}(k_BT)^{3/2}.\label{eqnew1}
\end{eqnarray}
The solution for Eqs. (\ref{eqnew1}) approximates
\begin{subequations}\label{solution}
\begin{eqnarray}
x&\approx& 1-\frac{8}{3\sqrt{\pi}}\alpha_s^{1/2}-\frac{m^{3/2}\zeta(3/2)}{2\sqrt{2}\pi^{3/2}\hbar^3n_0}(k_BT)^{3/2}+\frac{\sqrt{2}m^{3/2}\zeta(3/2)}{\pi^2\hbar^3n_0}(k_BT)^{3/2}\alpha_s^{1/2},\label{x}\\
{\cal M}&\approx&2-\frac{64}{3\sqrt{\pi}}\alpha_s^{1/2}+\frac{8\sqrt{2}m^{3/2}\zeta(3/2)}{\pi^2\hbar^3n_0}(k_BT)^{3/2}-\frac{4m^3\zeta(3/2)^2}{3\pi^{7/2}\hbar^6n_0^2}(k_BT)^{3}\alpha_s^{1/2}\nonumber\\
&&-\frac{\sqrt{2}m^{3/2}\zeta(3/2)}{\pi^{3/2}\hbar^3n_0}(k_BT)^{3/2}+\frac{m^3\zeta(3/2)^2}{4\pi^3\hbar^6n_0^2}(k_BT)^{3}.\label{M}
\end{eqnarray}\label{Mx}
\end{subequations}
At the critical temperature, the order parameter vanishes. Therefore, Eq. (\ref{x}) gives
\begin{eqnarray}
T_C\approx \frac{2\pi\hbar^2}{mk_B}\left[\frac{n_0}{\zeta(3/2)}\right]^{2/3}+\frac{16\sqrt{\pi}\hbar^2}{9mk_B}\left[\frac{n_0}{\zeta(3/2)}\right]^{2/3}\alpha_s^{1/2}.\label{Tc}
\end{eqnarray}
It is interesting to see that, for the ideal Bose gas, the gas parameter is zero so that Eq. (\ref{Tc}) remains only the first term in right-hand side, which exactly coincides with (\ref{Tideal}).
As a illustration, we consider the temperature-dependence of the order-parameter, or in other word, the temperature-dependence of the condensate density in the first experiment created BEC of Anderson {\it et. al.} \cite{Anderson1995}. In this experiment, isotope Rubidium 87 ($^{87}$Rb) was confined by magnetic fields and evaporatively cooled. The parameter for the $^{87}$Rb is mass of $m=86.909u$ with $u=1.66053873\times 10^{-27}$ kg atomic mass unit. The condensate BEC appeared near temperature of 170 nK at density $2.5\times 10^{12}~{\rm cm}^{-3}$. The reduced temperature-dependence of the condensate fraction associated with the isotope $^{87}$Rb in vicinity of the critical point is pictured graphically in Fig. \ref{fig:f1} by the solid line. It decreases as temperature increases and vanish at $T/T_C=1$. The dashed line is plotted for the corresponding ideal Bose gas. It is evident that this is second order phase transition.

It is also very conspicuous from Eq. (\ref{Tc}) since the condensed fraction (\ref{x}) can be rewritten in form
\begin{eqnarray}
x\approx\left(1-\frac{8\alpha_s^{1/2}}{3\sqrt{\pi}}\right)\left[1-\left(\frac{T}{T_C}\right)^{3/2}\right].\label{x1}
\end{eqnarray}
It is obvious that in terms of the gas parameter and reduced temperature, the condensed fraction Eq. (\ref{x1}) is universal and it reduces to that for the homogeneous ideal Bose gas whenever the gas parameter vanishes $\alpha_s=0$ \cite{Pitaevskii}.

Let us now work on the other main purpose of present paper, which is the effect from the interatomic interaction on the critical temperature, in particularly the shift of the critical point. It is easy to see that the repulsive interatomic interaction increases the critical temperature amount that equals the second term in right-hand side of Eq. (\ref{Tc}). The relative shift can be readily read
\begin{eqnarray}
\frac{\Delta T_C}{T_C^{(0)}}=c.\alpha_s^{1/3}.\label{shift1}
\end{eqnarray}
Evidently, Eq. (\ref{shift1}) has the same form with (\ref{shift}) and value of constant $c=8/9\sqrt{\pi}\approx 0.501$. As mentioned above, the sign and value of the constant $c$ is still controversial. In the early stage, it was found to be nagative. The typical result belongs to Martin {\it et. al.} in Ref. \cite{Wilkens2000}, in which the authors used perturbative solution of the crossover equations in the canonical ensemble and found $c=-0.93$. The negative sign of constant $c$ was also predicted in Ref. \cite{Fetter1971} within Hartree-Fock theory and Ref. \cite{Toyoda1982} by using renormalization group calculation.  Recently, most relevant studies have predicted that the constant $c$ is positive.  The reasonable physical arguments for positive $c$ was proposed by Gr\"{u}eter {\it et.al.}  \cite{Grueter1997} in microscopic scale. Accordingly, the repulsive interatomic interaction makes every atom more easily find a neighbor at a suitable distance (is order of the thermal de Broglie wavelength) to condensate. This implies that the critical temperature of the homogeneous weakly interacting Bose gas is higher than that of corresponding ideal Bose gas. Nonetheless, the order of magnitude of the constant $c$ varies in a large range. It was reported $c=0.34$ \cite{Grueter1997}, $c=0.7$ \cite{Holzmann1999}. Our result $c=0.501$ the average of that in Refs. \cite{Grueter1997,Holzmann1999}. By far, there are many authors who calculated the shift of critical temperature of the dilute Bose gas. Several typical results are listed in Table \ref{tab1}. The common character can be conspicuously recognized is that the relative shift of the critical temperature rises linearly with the scattering length at fixed particle density.

\begin{table}
\caption{\label{tab1}Value of constant $c$ in Eq. (\ref{shift1}) and critical temperature of the isotope potassium 39.}
\begin{ruledtabular}
\begin{tabular}{rl}
$c$ & Method\\
\hline
--3.524 & Renormalization group calculation \cite{Toyoda1982}\\
-- 0.93& Perturbative solution of the crossover equations \cite{Wilkens2000}\\
0.34& Path-integral Monte Carlo simulations and finite-size scaling \cite{Grueter1997}\\
0.501 & CJT effective action approach (present work)\\
0.7 & Ursell-Dyson approximation \cite{Holzmann1999}\\
1.23 & Exact renormalization group \cite{Ledowski2004}\\
1.3 & Simulations of classical field theory \cite{Davis2003}\\
1.32& Lattice simulation of O(2)$\varphi^4$ theory \cite{Arnold2001}\\
2.33 & The large-$N$ expansion at order $1/N$ \cite{Baym2000}\\
3.059 & Nonperturbative optimized linear $\delta$ expansion method \cite{SouzaCruz2001}\\
4.76 & Functional formulation of the Keldysh theory \cite{Stoof1992}
\end{tabular}
\end{ruledtabular}
\end{table}

Now we consider the non-condensate fraction of the dilute BEC at finite temperature. Inverting Eq. (\ref{chemical}) one easily see that the condensate density in the IHF approximation from differentiating the pressure with respect to the chemical potential, i.e.,
\begin{equation}
n=\frac{\partial {\cal P}}{\partial \mu}.\label{eq:rho}
\end{equation}
Combining equations \eqref{eq:press}-\eqref{eq:rho}, the condensate density is expressed in terms of the order parameter $\psi_0$ and the momentum integrals $P_{11}$ and $P_{22}$ \cite{VanThu2022},
\begin{equation}
n=\psi_0^2+\frac{1}{2}(P_{11}+P_{22}).\label{eq:n}
\end{equation}
The density of the depletion is defined as the number of particles in excited states per unit volume \cite{Pethick}. Based on equation \eqref{eq:n} one can easily see that the non-condensate fraction for a homogeneous dilute Bose gas is
\begin{equation}
n_{\rm ex}=\frac{1}{2}(P_{11}+P_{22}).\label{eq:ne}
\end{equation}
Substituting Eqs. (\ref{M}) into (\ref{chottichphan}) and then (\ref{eq:ne}) one has
\begin{eqnarray}
\frac{n_{\rm ex}}{n_0}\approx\frac{8}{3\sqrt{\pi}}\alpha_s^{1/2}+\frac{m^{3/2}\zeta(3/2)}{2\sqrt{2}\pi^{3/2}\hbar^3n_0}(k_BT)^{3/2}
-\frac{\sqrt{2}m^{3/2}\zeta(3/2)}{\pi^2\hbar^3n_0}(k_BT)^{3/2}\alpha_s^{1/2}.\label{nex}
\end{eqnarray}
It is easily seen that the non-condensate fraction (\ref{nex}) is exact the last three terms in right-hand side of the condensate fraction (\ref{x}) with the opposite sign. This fact reflects that the total of the condensate and non-condensate fractions equal the total particle density. Some important things can be read of from (\ref{nex}): (i) At zero temperature, right-hand side only remains the first term, which associated with the quantum fluctuations on top of the ground state. This term was first discovered by Bogoliubov \cite{Bogolyubov} and later by many authors \cite{Bogolyubov,Stringari,Carlen}, particularly was confirmed in the experiment \cite{Lopes}; (ii) in case of the ideal Bose gas, the gas parameter vanishes, right-hand side only contains the second term, which the thermal fluctuations of the ideal Bose gas at finite temperature. This result is in agreement with that of other authors [cf. Eq. (2.30) in Ref. \cite{Pethick}]. It is also shown that at zero temperature, quantum fluctuations do not appear in the ideal Bose gas \cite{Thu}; (iii) The non-condensate fraction of the dilute weakly interacting Bose gas contains an interferential term, which includes both thermal excitations and the interaction-induced quantum fluctuations. It is not difficult to check that at $T=T_C$ determined by Eq. (\ref{Tc}) the effective mass (\ref{M}) vanishes whereas the non-condensate fraction (\ref{nex}) approaches unity. This implies that all atoms are in excited states and the condensation disappears.

We next calculate the specific heat at constant volume, which is defined as product of temperature with first derivative of entropy $s$ with respect to the temperature \cite{Pethick},
\begin{eqnarray}
C_V=T\frac{\partial s}{\partial T},\label{cv}
\end{eqnarray}
in which the entropy is first derivative of the pressure with respect to temperature
\begin{eqnarray}
s=\frac{\partial {\cal P}}{\partial T}.\label{s}
\end{eqnarray}
To investigate the specific heat capacity, we consider two different regions of temperature, which are separated by the critical temperature. From Eqs. (\ref{eq:press}), (\ref{eq:n}), the pressure can be expressed in terms of the condensate density
\begin{equation}
{\cal P}=\frac{g}{2}n^2+gnP_{11}-\frac{g}{2}P_{11}^2-\frac{1}{2}\int_T\mbox{tr}\ln D^{-1}_{\rm{(IHF)}}(k).\label{eq:p1}
\end{equation}
The rule for the summation of Matsubara frequencies \cite{Schmitt},
\begin{equation}
k_BT\sum_{n=-\infty}^{n=+\infty}\ln\left[\omega_n^2+E^2(k)\right]=E(k)+2k_BT\ln\left[1-e^{-E(k)/k_BT}\right],
\end{equation}
allows us to write the last term in right-hand side of (\ref{eq:p1}) as
\begin{eqnarray}
\frac{1}{2}\int_T\mbox{tr}\ln D^{-1}_{\rm{(IHF)}}(k)=\int\frac{d^3k}{(2\pi)^3}E(k)+k_BT\int\frac{d^3k}{(2\pi)^3}\ln\left[1-e^{-E(k)/k_BT}\right].\label{e1}
\end{eqnarray}
Similarly to what we did for the momentum integrals, at high temperature and just below the critical temperature (\ref{e1}) can be approximated
\begin{eqnarray}
\frac{1}{2}\int_T\mbox{tr}\ln D^{-1}_{\text{(IHF)}}(k)\approx \frac{(2m)^{3/2}M^{5/2}}{30\pi^2\hbar^3}-\frac{2\pi\hbar^2\zeta(5/2)}{m\lambda_B^5}.\label{e2}
\end{eqnarray}
Based on Eqs. (\ref{chottichphan}), (\ref{Mx}) and (\ref{e1}), the pressure (\ref{eq:p1}) can be found to leading term of the gas parameter, or equivalently, to first order term of the scattering length
\begin{eqnarray}
P\approx\frac{2\pi\hbar^2n_0^2a_s}{m}+\frac{2\pi\hbar^2\zeta(5/2)}{m\lambda_B^5}+\frac{2\pi\hbar^2\zeta(3/2)^2a_s}{m\lambda_B^6}.\label{below}
\end{eqnarray}
One can easily to see that the first term in right-hand side of Eq. (\ref{below}) is the bulk pressure $P_0=gn_0^2/2$, the second term is the pure thermal pressure and the last one is contributed from both the interatomic interaction and thermal paressure. Eqs. (\ref{cv}), (\ref{s}) and (\ref{below}) give the dimensionless specific heat capacity for region $T<T_C$
\begin{eqnarray}
\frac{C_V}{n_0k_B}\approx\frac{15}{4}\frac{\zeta(5/2)}{n_0\lambda_B^3}+\frac{6\zeta(3/2)^2a_s}{n_0\lambda_B^4}.\label{CVtren}
\end{eqnarray}
\begin{figure}
  \includegraphics[width = 0.6\linewidth]{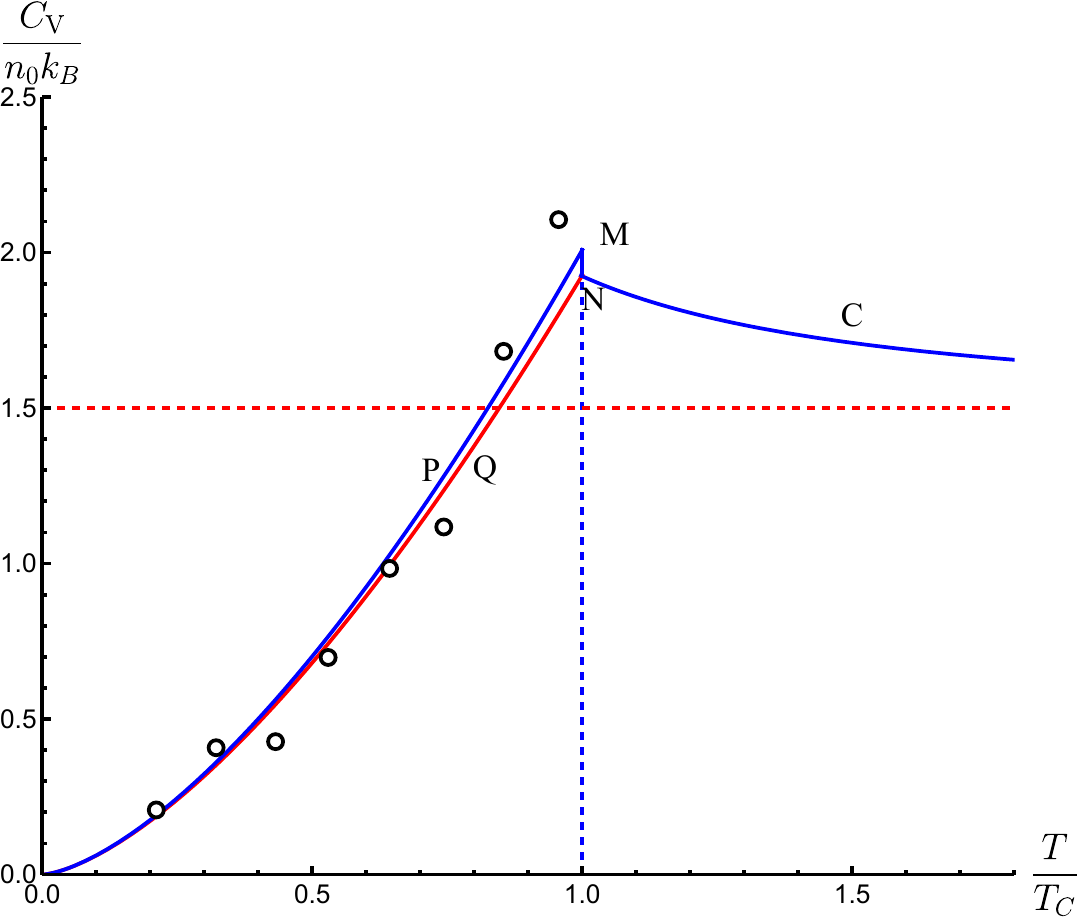}
\caption{(Color online) The reduced specific heat at constant volume of Bose gas of isotope $^{87}$Rb as a function of reduced temperature $T/T_C$ in vicinity of the critical temperature. The curves PMC and QNC correspond to $s$-wave scattering length $a_s=100a_0$ and 0, respectively. The open circles are the experimental data \cite{Shiozaki2014}.}
\label{fig:f2}       
\end{figure}

Let us move to the region $T>T_C$, in which all atoms are in excited states and the condensate fraction is zero and the system is in the normal phase. In this region, the CJT effective potential (\ref{eq:VIHF}) has only one minimum at $\psi_0^2=0$. The dispersion relation (\ref{disperIHF}) is modified in form
\begin{eqnarray}
E(k)=\frac{\hbar^2k^2}{2m}-\bar\mu,\label{dispersionnew}
\end{eqnarray}
where the effective chemical potential is defined as
\begin{eqnarray}
\bar\mu=\mu-\Sigma_2.\label{chemicaleffect}
\end{eqnarray}
The explicitly-independent parts of the temperature in momentum integrals (\ref{tichphan}) are divergent. This problem can be dealt by counter terms and they do not have any contribution on results. The temperature-dependent parts of the momentum integrals are in form of the Bose-integrals and thermal de Broglie wave length \cite{Olver2010},
\begin{eqnarray}
P_{11}=P_{22}=\frac{4}{\sqrt{\pi}\lambda_B^3}{\rm Li}_{3/2}(\bar z),\label{P1tren}
\end{eqnarray}
in which $\bar z=e^{\bar\mu/k_BT}$ is known as the fugacity of Bose gas and the polylogarithmic function is defined as
\begin{eqnarray}
{\rm Li}_n(x)=\sum_{k=1}^\infty \frac{x^k}{n^n}.\label{Li}
\end{eqnarray}
Similarly, owing to vanishing condensate, the classical and zero point terms in (\ref{e1}) do not contribute to result, therefore one has
\begin{eqnarray}
\frac{1}{2}\int_T{\rm Tr}\ln D^{-1}(k)=-\frac{2\pi\hbar^2}{m\lambda_B^5}{\rm Li}_{5/2}(\bar z).\label{integraltren}
\end{eqnarray}
The pressure in the normal phase can be easily derived
\begin{eqnarray}
{\cal P}=-\frac{1}{2}\int{\rm Tr}\ln D^{-1}(k)+gP_{11}^2.\label{Ptren}
\end{eqnarray}
As was pointed out in Ref. \cite{Haugset1998}, within this accuracy we may set $\bar\mu=\mu$ in the self-energy near the critical temperature. Consequently, Eqs. (\ref{P1tren}), (\ref{integraltren}) allow us to write the pressure (\ref{Ptren}) in form
\begin{eqnarray}
{\cal P}=\frac{2\pi\hbar^2}{m\lambda_B^5}{\rm Li}_{5/2}(\bar z)+\frac{64\hbar^2a_s}{m\lambda_B^6}{\rm Li}_{3/2}^2(\bar z).\label{Ptren1}
\end{eqnarray}
In terms of the particle density, Eq. (\ref{Ptren1}) gives
\begin{eqnarray}
n_0=\frac{64a_s}{\pi\lambda_B^4}{\rm Li}_{1/2}(\bar z){\rm Li}_{3/2}(\bar z)+\frac{1}{\lambda_B^3}{\rm Li}_{3/2}(\bar z).\label{dens}
\end{eqnarray}
Solving equation (\ref{dens}) one can expresses the polylogarithmic function in form
\begin{eqnarray}
{\rm Li}_{3/2}(\bar z)=n_0\lambda_B^3\left[1+\Lambda\left(-1-\frac{\pi\ln\bar z}{64n_0a_s\lambda_B^2}\right)\right],\label{dens1}
\end{eqnarray}
in which $\Lambda(x)$ is the Lambert function \cite{Olver2010}. For the dilute weakly interacting Bose gas at finite temperature, not only the gas parameter is very small but also the single-particle excitations prevail, or equivalently, the de Broglie wavelength is much smaller than the healing length $n_0a_s\lambda_B^2\ll1$ \cite{Pethick}. Therefore, at the lowest order approximation, the effective chemical potential can be attained from (\ref{dens1}),
\begin{eqnarray}
\bar\mu=k_BT\ln{\rm Li}^{-1}_{3/2}(n_0\lambda_B^3),\label{mu}
\end{eqnarray}
or equivalently
\begin{eqnarray}
{\rm Li}_{3/2}(\bar z)=n_0\lambda_B^3.\label{Li32}
\end{eqnarray}
Using the relationship of the polylogarithm function
\begin{eqnarray}
\frac{d{\rm Li}_{n}(\bar z)}{d\bar z}=\frac{1}{\bar z}{\rm Li}_{n-1}(\bar z),\label{dhLi}
\end{eqnarray}
together with Eq. (\ref{Li32}) one has
\begin{eqnarray}
{\rm Li}_{5/2}(\bar z)=\frac{\lambda_B^3{\cal P}}{k_BT},\label{Li52}
\end{eqnarray}
in which ${\cal P}=n_0[\bar\mu+{\cal C}]$ with constant ${\cal C}$ ensuring the continuity of the pressure at the critical temperature. Eqs. (\ref{Li32}) and (\ref{Li52}) exactly coincide with the equations of state of the ideal Bose gas \cite{Huang1987}, in which the effective chemical potential of the dilute weakly interacting Bose gas plays the role of the one associated with the ideal Bose gas. As pointed out in many textbooks on BEC \cite{Pethick,Huang1987,Pitaevskii}, above the critical temperature, the specific heat capacity at constant volume can be expressed as
\begin{eqnarray}
\frac{C_V}{n_0k_B}=\frac{15}{4}\frac{{\rm Li}_{5/2}(\bar z)}{n_0\lambda_B^3}-\frac{9}{4}\frac{{\rm Li}_{3/2}(\bar z)}{{\rm Li}_{1/2}(\bar z)}.\label{above}
\end{eqnarray}
To complete these calculations, we now evaluate the effective chemical potential in the normal phase. From the definition (\ref{chemicaleffect}) one has
\begin{eqnarray}
\bar \mu=\mu-2gP_{11},\label{effect1}
\end{eqnarray}
with $P_{11}$ given in (\ref{P1tren}).
The behaviour of the reduced specific heat capacity in vicinity of the critical temperature of the isotope $^{87}$Rb is graphically sketched in Fig. \ref{fig:f2} by curve PMC for the same value of parameters as in Fig. \ref{fig:f1}. The curve has the typical behaviour, in particular, it shows a "peak" with a finite jump at the critical temperature and it is not difficult to recognize the well-known behaviour of $\lambda$-shape. In the classical limit, the specific heat capacity tends to 3/2. Below the crtitical temperature, the experimental data of the specific heat are marked by open circles \cite{Shiozaki2014}. It is worth to note that in this experiment \cite{RomeroRochin2012}, the BEC of the isotope $^{87}$Rb was produced in a double magneto-optical trap, in which trapping potential is harmonic with
frequencies $\omega_x=\omega_0,~ \omega_y=\omega_z=9\omega_0$, where $\omega_0=2\pi\times 23$ Hz. In that case, the size of BEC cloud $r=\sqrt{k_BT/m\omega^2}$ with $\omega^3=\omega_x\omega_y\omega_z$. This value is much larger than the healing length therefore the effect of the finite-size can be neglected. The excellent agreement between our result and the experimental data is showed.

 It is also interesting to check from Eqs. (\ref{above}) and (\ref{effect1}) that at a given value of the $T/T_C$, the interatomic interaction do not change value of the specific heat and it purely shifts the critical temperature in compared with those of the ideal Bose gas as was proved in \cite{Kim2007}. In case of the ideal Bose gas, the scattering length vahishes. In region above the critical temperature, Eq. (\ref{above}) does not change whereas the right-hand side of Eq. (\ref{CVtren}) remains only the first term. We obtain the well-known result for the specific heat capacity of the ideal Bose gas, which is now showed by curve QNC with typical characteristics \cite{Pethick,Kim2007,AlSugheir2023,Huang1987}:

- The secific heat capacity is continuous at the critical temperature and, of course, its first derivative with respect to temperature will be discontinuous at this point.

- In condensate phase, the specific heat capacity is proportional to the temperature in law $C_V/n_0k_B\propto T^{3/2}$.

Finally, it is of interest to investigate the critical exponents of the dilute weakly interacting Bose gas in our model. To do this, we define the reduced temperature as
\begin{eqnarray}
t=\frac{\left|T-T_C\right|}{T_C}.\label{t}
\end{eqnarray}
In vicinity of the critical temperature, the reduced temperature is very small, i.e. $t\ll1$. Substituting (\ref{t}) into the order parameter (\ref{x}) yields
\begin{eqnarray}
x\approx \left(\frac{3}{2}-\frac{4}{\sqrt{\pi}}\alpha_s^{1/2}\right)t.\label{t1}
\end{eqnarray}
It is obvious that the critical exponent for the order parameter is $\beta=1$ (do not confuse with the inverse temperature). This result was also obtained by other authors \cite{Fisher1967,ReyesAyala2019,Ma1985} and it once again confirms the second order of phase transition in dilute Bose gas.
The diffusion Monte Carlo simulations \cite{Giorgini1999} pointed out that the diluteness condition is quantitatively satisfied $\alpha_s<10^{-3}$. This means that the second term in right-hand side of Eq. (\ref{t1}) is negligible and this equation becomes $x\approx 3t/2$. This implies that this critical exponent is the same for all dilute homogeneous weakly interacting Bose gas and thus it is universal.

We next find the critical exponent $\alpha$, which characterizes the behaviour of the specific heat capacity in vicinity of the critical point. The existence of the "peak" at the critical point suggests us define this critical exponent that turns out to be negative. Above the critical temperature, the specific heat capacity is written as \cite{ReyesAyala2019},
\begin{eqnarray}
\frac{C_V}{n_0k_B}\approx \frac{15}{4}\zeta(5/2)+\frac{9}{16}\left[5\zeta(5/2)-\frac{9\zeta(3/2)^3}{\pi}\right]t.\label{criticalcv1}
\end{eqnarray}
Likewise, below critical temperature one finds
\begin{eqnarray}
\frac{C_V}{n_0k_B}\approx\frac{15\zeta(5/2)}{3\zeta(3/2)}+6\zeta(3/2)^{2/3}\alpha_s^{1/3}-\left[\frac{45\zeta(5/2)}{8\zeta(3/2)}+12\zeta(3/2)^{2/3}\alpha_s^{1/3}\right]t.\label{criticalcv2}
\end{eqnarray}
Eqs. (\ref{criticalcv1}) and (\ref{criticalcv2}) allow us to identify the critical exponent $C_V\propto |t|^{-\alpha}$ \cite{Fisher1967,Ma1985} as given by $\alpha=-1$ \cite{ReyesAyala2019}.
The other critical exponents for the dilute homogeneous weakly interacting Bose gas can be obtained by means
of equalities among those exponents. The first is the scaling hypothesis, introduced by
Widom \cite{Fisher1967,Ma1985,Widom1965} and the later was validated by the renormalization group theory developed by Wilson \cite{Wilson1974}. These equalities are called Josephson law
\begin{eqnarray}
\nu d=2-\alpha,\label{Josephson}
\end{eqnarray}
Rushbrooke law
\begin{eqnarray}
\alpha+2\beta+\gamma=2,\label{Rushbrook}
\end{eqnarray}
Widom law
\begin{eqnarray}
\gamma=\beta(\delta-1),\label{Widom}
\end{eqnarray}
and Fisher law
\begin{eqnarray}
\gamma=\nu(2-\eta).\label{Fisher}
\end{eqnarray}
From these laws one finds $\gamma=1,\delta=1,\nu=1$ and $\eta=1$. It is obvious that this is a critical universality class on its own. This fact is valid not only for the ideal Bose gas but also for the homogeneous dilute weakly interacting Bose gas.

\section{Conclusion and outlook\label{sec:4}}

Beyond the standard Bolgoliubov theory, we have explored the CJT effective action approach to investigate the thermodynamic properties at finite temperature of homogeneous dilute interacting Bose gas. After deriving the CJT effective potential in the IHF approximation, where the Goldstone theorem is satisfied, the equations of state are obtained. Based on these equations, the important properties of the dilute Bose gas, namely, the critical temperature, non-condensate fraction, specific heat capacity, critical exponents have been studied. Two main results are in order:

- The critical temperature is found in two-terms: one corresponds to that of the ideal Bose gas and the other depends on the strength of atomic interaction. The presence of the atomic interaction makes a shift in the critical temperature in compared with the one of the ideal Bose gas. This shift can be expressed in the universal form (\ref{shift}). It is worthy of interest to write (\ref{shift}) in a more general form
\begin{eqnarray}
\frac{\Delta T_C}{T_C^{(0)}}=c.\alpha_s^{a}.\label{shift2}
\end{eqnarray}
The two constants $c$ and $a$ determine the behaviors of the shift of critical temperature of Bose gas at low density. One of the outstanding achievements is that most of works found $a=1/3$ as listed in Table \ref{tab1}, some other values were attained, for example, $a=1/2$ in Ref. \cite{Huang1964}, $a=0.34\pm0.03$ in Ref. \cite{Grueter1997}. It is worth to know that the CJT effective action approach was employed by Haugset {\it et al.} to calculate the critical temperature of the dilute weakly interacting Bose gas in one-loop approximation \cite{Haugset1998}. Their result showed that the critical temperature is the same as the one of the ideal Bose gas. This implies that the contribution from the two-loop diagrams are significantly important. Taking into account the higher-order terms of the gas parameter, our result can be achieved
\begin{eqnarray}
\frac{\Delta T_C}{T_C^{(0)}}=\left(\frac{8}{9\sqrt{\pi}}+\frac{272}{81\pi}\alpha_s^{1/2}\right)\alpha_s^{1/2}.\label{high}
\end{eqnarray}
The condition of diluteness shows that the contribution from the higher-order terms of the gas parameter is negligible on the shift of the critical temperature.

- The non-condensate fraction can be separated into three terms: first one is pure quantum fluctuations, second one described the thermal fluctuations and the last one includes both the quantum and thermal fluctuations. An important thing should be noted is the difference between our result (\ref{nex}) and (\ref{Shi}). In Eq. (\ref{nex}), the non-condensate fraction depends on temperature in half-integer power law instead of the integer one in Eq. (\ref{Shi}), which was first exhibited by Glassgold {\it et al.} \cite{Glassgold1960}. This half-integer power law of temperature-dependence of the condensate fraction was also found in Ref. \cite{CapogrossoSansone2010} within Beliaev’s diagrammatic technique and recently confirmed again in Ref. \cite{AlSugheir2023} by using  static fluctuation approximation. In addition, our result contains the interferential term whereas this term is absence in Eq. (\ref{Shi}).

- The specific heat capacity is investigated in both regions of temperature. Its step of jump at the critical temperature approximates
\begin{eqnarray}
\frac{\Delta C_V}{(n_0k_B)}\bigg|_{T=T_C}\approx 6\zeta(3/2)^{2/3}\alpha_s^{1/3}.
\end{eqnarray}
This can be reduced to the well-known result for the ideal Bose gas if the gas parameter vanishes.

- Two critical exponents associated with the order-parameter and specific heat capacity are found via the analytical calculations. Based on that, the remaining critical exponents are also calculated by using four equalities among them.

We would insist again that our system under consideration is investigated in the grand canonical ensemble thus the total particle density is fixed. This implies that our system is connected to a "reservoirs" of particles while it is confined by a potential, even in the state-of-the-art relevant experiments. Consequently, the number of particles is finite but the grand canonical ensemble is still used \cite{Politzer1996}.

It is also very interesting to explore these calculations for Bose-Einstein condensate mixtures. The interspecies interaction is expected to produce some novel results, especially to compare with relevant results in \cite{Shi2000}. Furthermore, this procedure can also be employed to investigate the condensed fraction of a Bose gas confined between two parallel plates and the resulting critical temperature and non-condensate fraction \cite{Thu2023}.

\begin{acknowledgements}
This research is funded by Vietnam National Foundation for Science and Technology Development (NAFOSTED) under grant number 103.01-2023.12. We are very grateful to G. D. Telles for the useful conversations and providing the experimental data.
\end{acknowledgements}

\section*{Conflict of interest}
All of the authors declare that we have no conflict of interest.

\bibliography{biblio.bib}

\end{document}